\documentclass[twocolumn,english,aps,pra,superscriptaddress]{revtex4-1}
\usepackage{graphicx}
\usepackage{amsmath}
\usepackage{amssymb}
\usepackage{braket}
\usepackage{bm}
\usepackage[colorlinks=true,linkcolor=blue,citecolor=blue,urlcolor=blue]{hyperref}
\usepackage{subfigure}
\usepackage{graphicx}
\usepackage{times}
\usepackage{mathptmx}

\makeatother

\usepackage{babel}
\begin{document}
\title{Magnetically Induced Optical Transparency With Ultra-Narrow Spectrum }
\author{Guohui Dong}
\affiliation{Beijing Computational Science Research Center, Beijing, 100084, China}
\affiliation{Graduate School of Chinese Academy of Engineering Physics, Beijing
100084, China}
\author{Dazhi Xu}
\email{dzxu@bit.edu.cn}

\affiliation{Center for Quantum Technology Research and School of Physics, Beijing
Institute of Technology, Beijing 100081, China}
\author{Peng Zhang}
\email{pengzhang@ruc.edu.cn}

\affiliation{Department of Physics, Renmin University of China, Beijing, 100872,
China}
\affiliation{Beijing Computational Science Research Center, Beijing, 100084, China}
\date{\today}
\begin{abstract}
Magnetically induced optical transparency (MIOT) is a technique to
realize the narrow transmission spectrum in a cavity quantum electric
dynamics (cavity QED) system, which is demonstrated in the recent
experiment of cold $^{88}$Sr atoms in an optical cavity [Phys. Rev.
Lett. \textbf{118}, 263601 (2017)]. In this experiment, MIOT induces a new narrow transmission window for the probe beam, which is highly immune to the fluctuation of the cavity mode frequency.
The linewidth of this transmission window approaches the decay rate of the electronic $^{3}{\rm P}_{1}$ state (about
$2\pi\times7.5$kHz) and is much less than the uncertainty of the
cavity mode frequency (about $2\pi\times150$kHz). In this work, we
propose an approach to further reduce the linewidth of this MIOT-induced
transmission window, with the help of two Raman beams which couples the electronic $^{3}{\rm P}_{1}$ state to the $^{3}{\rm S}_{1}$ state,
and the $^{3}{\rm S}_{1}$ state to the $^{3}{\rm P}_{0}$ state, respectively. With this approach, one can reduce the transmission linewidth  by orders of magnitude. Moreover, the peak value of the relative transmission power or the transmission rate of the probe beam is almost unchanged by the Raman beams.
Our results are helpful for the study of precision measurement and other quantum optical processes based on cavity quantum electronic dynamics (cavity-QED).
\end{abstract}
\maketitle

\section{Introduction}

In recent years, there has been a series of efforts to realize optical
systems with the narrow transmission spectrum. This type of system can
play crucial role in the precision measurements~\citep{Harry_AO2006, Eisele_PRL2009, Abramovici_Science1992, Abbott_RPP2009, Graham_PRL2013, Rosi_Nature2014}
and the improvement of frequency stability of laser beams~\citep{Drever_APB1983, Salomon_JOSAB1988, Young_PRL1999, Hinkley_Science2013, Bloom_Nature2014, Ushijima_NatPho2015}.
For the realization of these narrow spectrum systems, a critical problem
is to overcome the negative influences from the thermal fluctuations
of the reference cavity, which is also the main limitation of the
linewidths of state-of-the-art laser systems~\citep{Numata_PRL2004, Notcutt_PRA2006, Kessler_NatPho2012, Kimble_PRL2008}.
In addition to improving the manufacturing craft of the cavity system,
one hopeful strategy to solve this problem is making use of the spin-forbidden
transitions of alkaline-earth (like) atoms, e.g., the transitions
between $^{1}{\rm S}_{0}$ and $^{3}{\rm P}_{0,1,2}$ states, which
have extremely narrow linewidths. Based on this idea, people have developed
a serious of approaches, such as superradiant laser~\citep{Meiser_PRL2009, Chen_CSB2009, Bohnet_Nature2012, Norcia_PRX2016}
and cavity-assisted nonlinear spectroscopy~\citep{Martin_PRA2011, Christensen_PRA2015}.

In 2017, M.~N.~Winchester~\textit{et al.} experimentally demonstrated
a new approach to realize narrow spectrum with alkaline-earth atoms
in the cavity, which is called as magnetically induced optical transparency (MIOT)
\citep{Winchester_PRL2017}. In this experiment, an ensemble of
cold $^{88}$Sr atoms is trapped in a high-finesse cavity which resonates
with the transition between $^{1}{\rm S}_{0}$ and $^{3}{\rm P}_{1}$
states. Due to the selection rule of this system, there is a dark
mode or a special atomic state in the $^{3}{\rm P}_{1}$ manifold,
which is not directly coupled to the $^{1}{\rm S}_{0}$ state by the cavity
photons. Nevertheless, when a static magnetic field
is applied, this dark mode is mixed with the atom-photon dressed
states, and then induces a new transmission window for the probe beam propagating
through this cavity. This new transmission window is localized in
the middle of the two vacuum Rabi peaks and is robust for the thermal
fluctuations of the cavity. The linewidth of this window is
much less than the variance of the frequency of the cavity mode (about
$2\pi\times150$kHz) and can approach the natural linewidth $\gamma$ of the
$^{3}{\rm P}_{1}$ state (about $2\pi\times7.5$kHz).

\begin{figure}[tbp]
\begin{centering}
\includegraphics[width=8cm]{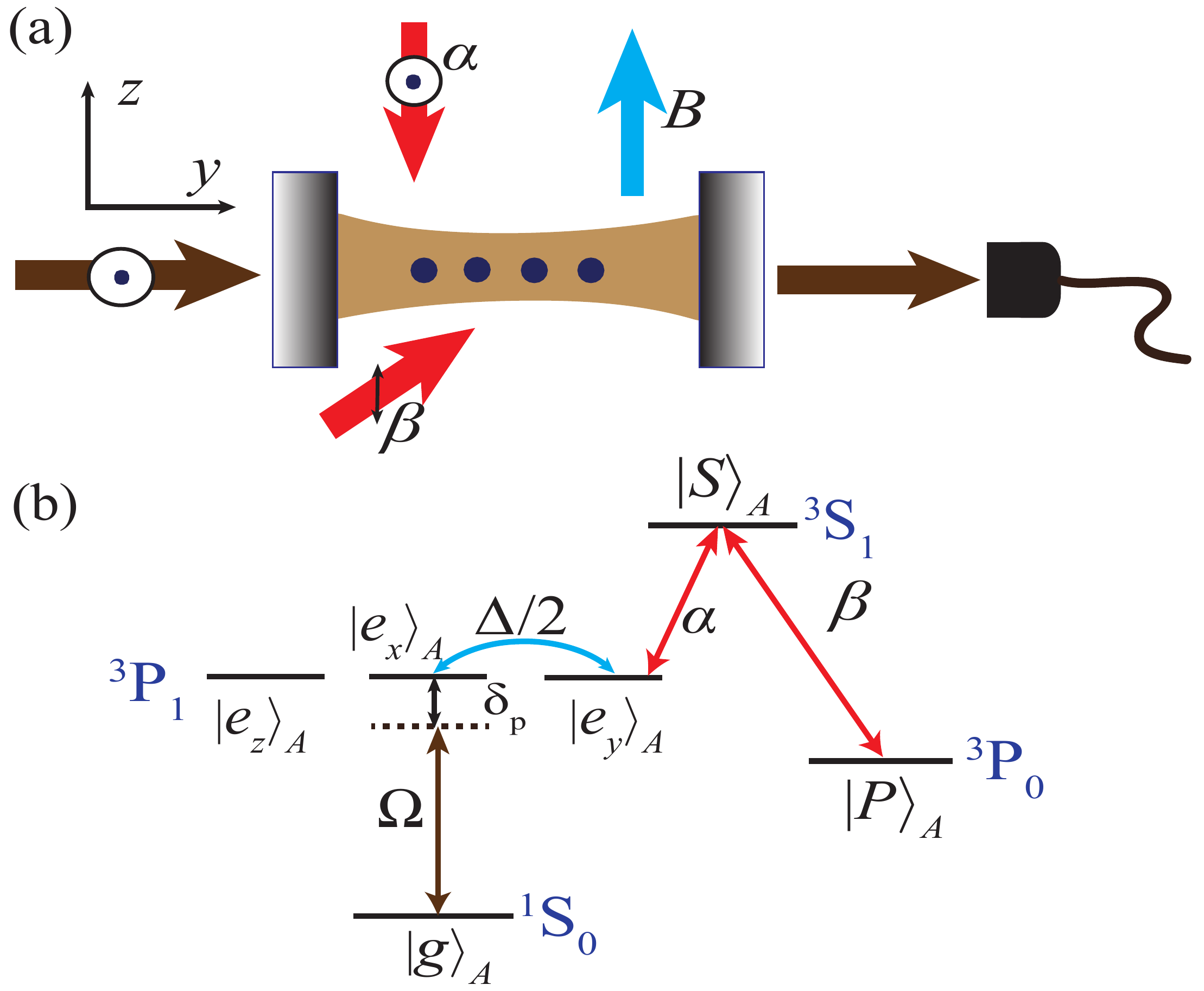}
%\par
\end{centering}
\caption{\label{fig:scheme1}(Color online) {\bf (a):} Schematic diagram of our proposal. The
bosonic alkaline-earth (like) atoms (e.g., $^{88}{\rm Sr}$ atoms)
atoms are trapped in an optical cavity. The MIOT effect is induced by a static magnetic field along the $z$-direction, and can be detected via the transmission spectrum of a probe beam through the cavity, which is polarized along the $x$- and propagating along the $y$-direction (i.e., the axial direction of the cavity). The modulation of the linewidth of the MIOT transmission peak is realized via
two Raman beams $\alpha$ and $\beta$, which
are polarized along the $x$-direction and $z$-direction, respectively.
{\bf (b):} The energy-level diagram of a
bosonic alkaline-earth (like) atom. Here
$\left|e_{x,y,z}\right\rangle_{A}$ are the $^{3}\mathrm{P}_{1}$ states defined in Eqs.~(\ref{exa}-\ref{eza}),
and  $\left|S\right\rangle _{A}$ is the $^{3}\mathrm{S}_{1}$ state with magnetic quantum number $m_J=0$. The cavity mode can couple
the $^{1}\mathrm{S}_{0}$ state $|g\rangle_A$ with $\left|e_{x}\right\rangle_{A}$,  the Raman beam $\alpha$ couples $\left|e_{y}\right\rangle_{A}$ and $\left|S\right\rangle_{A}$, and the Raman beam $\beta$ couples $\left|S\right\rangle_{A}$ and the $^{3}\mathrm{P}_{0}$ state $|P\rangle_A$.
In addition, the $^{3}\mathrm{P}_{1}$ states $\left|e_{x}\right\rangle_{A}$
and $\left|e_{y}\right\rangle_{A}$ can be coupled by the
the magnetic field via the Zeeman effect, with coupling intensity $\Delta/2$. Other notations are all defined in Sec.~II.}
\end{figure}

It is natural to ask whether we can further improve the
MIOT technique and decrease the linewidth. A straightforward idea
is to replace the $^{3}{\rm P}_{1}$ states with other atomic metastable
states with extremely small natural linewidth, e.g., the
$^{3}{\rm P}_{0}$ states (linewidth $\sim$mHz). However, the coupling
between the cavity mode and the $^{3}{\rm P}_{0}$ states is too weak. As
a result, the MIOT-induced transmission peak and the two vacuum
Rabi peaks would merge as a single peak, with the linewidth being
as large as the variance of the cavity mode frequency. Thus, this direct scheme does not work.

In this paper, we propose an approach to effectively reduce
the linewidth of the MIOT transmission peak, while keeping
this peak clearly distinguished from the other two vacuum Rabi
peaks.  Our basic idea is to apply two Raman
beams $\alpha$ and $\beta$ to the alkaline-earth atom, which couple the electronic
$^{3}{\rm P}_{1}$ states with $^{3}{\rm S}_{1}$ state and $^{3}{\rm S}_{1}$
state with $^{3}{\rm P}_{0}$ states, respectively (Fig.~\ref{fig:scheme1}).
As shown in Sec.~II, by choosing particular polarizations for these beams,
one can make the beam $\alpha$ only couple the original dark mode in the $^{3}{\rm P}_{1}$ manifold
to the $^{3}{\rm S}_{1}$ state while not influencing other relevant $^{3}{\rm P}_{1}$ states.
As a result, in the presence of two Raman beams,
the dark mode of our system becomes a dressed state of $^{3}{\rm P}_{1}$
and $^{3}{\rm P}_{0}$ levels, rather than a pure $^{3}{\rm P}_{1}$
state. Since the natural linewidth of $^{3}{\rm P}_{0}$
state is very small (of the order of mHz), the linewidth $\gamma_D$ of the new
dark mode can also be manipulated to be much smaller than the natural linewidth $\gamma$ of the
$^{3}{\rm P}_{1}$ state. Therefore, when the magnetic field is
applied, the linewidth of the transmission peak induced by the MIOT
effect, which can approach $\gamma_D$, is able to be reduced by orders of magnitude from $\gamma$. Meanwhile, the locations of this transmission peak and the vacuum Rabi peaks
are almost not changed by the Raman beams, and thus the former can
still be clearly distinguished from the latter ones. We also show that the hight of this MIOT-induced peak (i.e., the
maximum transmission rate) is also almost unchanged by the Raman beams, and the position of this  peak is
robust for the cavity frequency and  the total frequency of the Raman beams. The uncertainty of this peak position is approximately same as the fluctuation of the frequency difference of the two Raman beams. As a new approach for the realization of ultra-narrow spectrum,
our results are helpful for the study of quantum optics and
precision measurement technique via cavity quantum electronic dynamics (cavity-QED).

The remainder of this paper is organized as follows. In Sec.~II we
show  the physical principle of our proposal. In Sec.~III we explicitly
derive the width, location, height and pulling coefficient
for the MIOT-induced transmission peak via  the Heisenberg-Langevin equation
approach. A summary of this work is given in Sec.~IV.

\section{Physical Principle of MIOT with ultra narrow linewidth\label{sec:Physicsal-Principle}}

\subsection{System and setup}

In our system, there is an ensemble of cooled bosonic alkaline-earth (like) atoms (e.g., $^{88}{\rm Sr}$ atoms) trapped in the optical cavity (Fig.~\ref{fig:scheme1}). In the experiments of Ref.~\citep{Winchester_PRL2017}, the atom
number $N$ is about $10^{6}$. In this section, for simplicity, we take the single-atom case (i.e., $N=1$)
as an example to illustrate the physical principle of our proposal. The explicit calculations for the multi-atom case are given in Sec.~III.

For the convenience of our discussion, we denote the $^{1}{\rm S}_{0}$ state as $|g\rangle_{A}$ and define three $^{3}{\rm P}_{1}$ states $|e_{x,y,z}\rangle_{A}$ as
\begin{eqnarray}
|e_{x}\rangle_{A} & = & -\frac{1}{\sqrt{2}}\left(|e_{\mathit{+}1}\rangle_{A}-|e_{\mathit{-}1}\rangle_{A}\right);\label{exa}\\
|e_{y}\rangle_{A} & = & \frac{i}{\sqrt{2}}\left(|e_{+1}\rangle_{A}+|e_{-1}\rangle_{A}\right);\label{eya}\\
|e_{z}\rangle_{A} & = & |e_{0}\rangle_{A},\label{eza}
\end{eqnarray}
where $|e_{m_{J}}\rangle_{A}$ ($m_{J}=0,\pm1$) is
the $^{3}{\rm P}_{1}$ state with magnetic quantum number along the $z$-axis being $m_{J}$.
As shown in Fig.~\ref{fig:scheme1}(a), the optical axis  of the
cavity is along the $y$-axis, so that the polarization of the photons in
the cavity are in the $x-z$ plane. As a result, the cavity photons only couple the atomic state $|g\rangle_{A}$
to $|e_{x,z}\rangle_{A}$, and cannot induce the transition
from $|g\rangle_{A}$ to $|e_{y}\rangle_{A}$ [Fig.~\ref{fig:scheme1}(b)].
The MIOT effect of this system is induced by a bias magnetic field $B$ along the $z$-direction, and can be detected via the transmission spectrum of an $x$-polarized pump laser transmitted through the cavity.

Above is this setup of the MIOT experiment in Ref.~\citep{Winchester_PRL2017}.
In the current proposal, we further assume two Raman laser beams
$\alpha$ and $\beta$ are applied [Fig.~\ref{fig:scheme1}(a)]. The beam $\alpha$ is polarized
along the $x$-axis and is resonant with the atomic transition $^{3}{\rm P}_{1}\leftrightarrow\:^{3}{\rm S}_{1}$ [Fig.~\ref{fig:scheme1}(b)].
According to the selection rule, this beam couples $|e_{y}\rangle_{A}$
to the $^{3}{\rm S}_{1}$ state with magnetic quantum number $m_{J}=0$
(state $|S\rangle_{A}$), and does not couple $|e_{x}\rangle_{A}$
to any $^{3}{\rm S}_{1}$ state (see Appendix~\ref{appnA}).
In addition, the Raman beam $\beta$ is polarized along the $z$-axis
and is resonant with the atomic transition between the state $|S\rangle_{A}$ and the $^{3}{\rm P}_{0}$ state (state $|P\rangle_{A}$) which has an extremely long lifetime.

It is also worth pointing out that  the cavity photon with $z$-polarization
and the atomic state $|e_{z}\rangle_{A}$ are totally decoupled from
other parts of our system, and are irrelevant for our problem. Thus,
in the following we only take into account the atomic states $|e_{x,y}\rangle_{A}$
and the photon with $x$-polarization.

\subsection{Positions and widths of transmission peaks}

For our problem, the atom and cavity field are initially prepared
in the state $|g\rangle_{A}|0\rangle_{c}$, where $|0\rangle_{c}$
is the vacuum state of the cavity. A weak probe laser beam polarized
along the $x$-axis with circular frequency $\omega_{p}$ is injected
into and transmitted through the cavity [Fig.~\ref{fig:scheme1}(a)].
The transmission spectrum is given by the intensity of the transmitted probe beam measured as a function of $\omega_{p}$. In the following two subsections, we will estimate the positions and widths of the peaks of this transmission spectrum for various cases. Before going into the detailed discussions, we first introduce our approach to these estimations.

For convenience, we denote $H$ as the Hamiltonian of the atom and cavity
field, which includes the coupling between the atom and the cavity
photon, the bias magnetic field, and the Raman beams, i.e., the self-Hamiltonian of our system. We further denote $H_{p}$ as the Hamiltonian which describes the coupling between the probe beam and the cavity field.
In our system, the probe beam is linearly coupled to the cavity field. Thus, we have ($\hbar=1$)
\begin{equation}
H_{p}=ig_p(e^{-i\omega_{p}t}\hat{a}_{x}^{\dagger}-\textrm{h.c.}),\label{hp}
\end{equation}
where $\hat{a}_{x}^{\dagger}$ is the creation operator of the cavity photon
polarized along the $x$-axis. The coupling intensity between the probe beam and the cavity photon is $g_p=\sqrt{\kappa I_{p}/2\omega_{p}}$ with
$I_{p}$ the probe-beam the intensity and
$\kappa$ the  photon dissipation rate of the cavity mode. Here we have assumed that the photon dissipation rate of the left and right cavity
mirrors are same.

For our system, the initial state $|g\rangle_{A}|0\rangle_{c}$ is
always the ground state of $H$, and we can choose its energy as $E_{0}=0$. As a result, the position and width of the peaks of the transmission spectrum can be estimated as follows:

(i) If $H_{p}$ can couple $|g\rangle_{A}|0\rangle_{c}$ with another
eigenstate $|\lambda\rangle$ of $H$ (i.e., $\langle\lambda|H_{p}|g\rangle_{A}|0\rangle_{c}\neq0$),
then a peak of the transmission spectrum can appear at
\begin{equation}
\omega_{p}=E_{\lambda},\label{con}
\end{equation}
where $E_{\lambda}$ is the eigenenergy of $H$ corresponding to $|\lambda\rangle$.

(ii) In our system, the excited states of $H$ can decay via the leaking
of the photon or the spontaneous emission of the $^{3}{\rm P}_{1}$
and $^{3}{\rm S}_{0}$ states. As a result, the eigenstate $|\lambda\rangle$ has a non-zero decay rate $\Gamma_{\lambda}$. The linewidth of the transmission peak at the location $\omega_{p}=E_{\lambda}$ can be estimated as this decay rate.

\subsection{Case with $B=0$: Rabi splitting}

We first consider the case without the magnetic field, i.e., $B=0$.
In this case the Hamiltonian $H$ of our system is given by
\begin{equation}
H=H_{AP}+H_{{\rm Raman}}.\label{hh}
\end{equation}
Here $H_{AP}$ is the Hamiltonian for the atomic $^{3}{\rm P}_{1}$
states and the cavity field, and is expressed as
\begin{eqnarray}
H_{AP} & = & \omega_{A}\hat{a}_{x}^{\dagger}\hat{a}_{x}+\omega_A\left(|e_{x}\rangle_{A}\langle e_{x}|+|e_{y}\rangle_{A}\langle e_{y}|\right)\nonumber \\
 &  & +\frac{\Omega}{2}\hat{a}_{x}|e_{x}\rangle_{A}\langle\mathrm{g}|+\textrm{h.c.},\label{hap}
\end{eqnarray}
where $\Omega$ is the Rabi frequency for the atom-photon
coupling, and  $\omega_A$ is
the atomic $^3{\rm P}_1\leftrightarrow{^1{\rm S}_0}$ transition frequency.
We assume the cavity mode is resonate with the atomic $^3{\rm P}_1\leftrightarrow{^1{\rm S}_0}$ transition as in the experiment of Ref.~\citep{Winchester_PRL2017}.
In addition, $H_{{\rm Raman}}$ in Eq. (\ref{hh}) is the Hamiltonian
for the atomic states $|S\rangle_{A}$, $|P\rangle_{A}$ and the Raman coupling, which is given by
\begin{eqnarray}
H_{{\rm Raman}} & = & \omega_{A}\left(|S\rangle_{A}\langle S|+|P\rangle_{A}\langle P|\right)\nonumber \\
 &  & +\frac{g_{\alpha}}{2}|S\rangle_{A}\langle e_{y}|+\frac{g_{\beta}}{2}|S\rangle_{A}\langle P|+\textrm{h.c.}.\label{hra}
\end{eqnarray}
Here $g_{\alpha(\beta)}$ is the Rabi frequency of the Raman beam $\alpha(\beta)$.
We have chosen a rotated frame to eliminate the time-dependence of the Hamiltonian.

\begin{figure}[tbp]
\begin{centering}
\includegraphics[width=8cm]{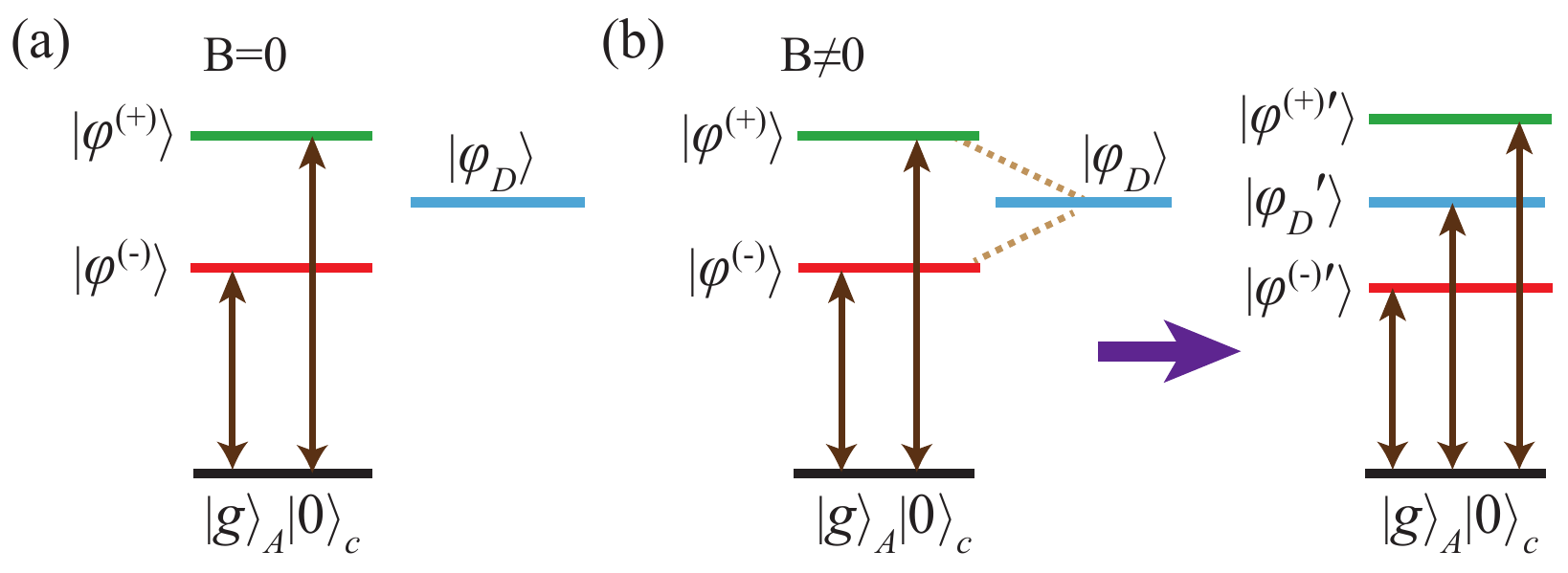}
\end{centering}
\caption{\label{fig:The-dressed-state-picture}(Color online)
The transitions induced by the probe beam. {\bf (a):} The case with $B=0$.
In this case, the probe beam only induces the transitions from the ground state
$|g\rangle_A|0\rangle_c$ to the two dressed states $|\varphi^{(\pm)}\rangle$, while does not couple
$|g\rangle_A|0\rangle_c$ to the dark mode $|\varphi_D\rangle$.
{\bf (b):} The case with $B\neq 0$. In this case, the dressed states $|\varphi^{(\pm)}\rangle$
and the dark mode $|\varphi_D\rangle$ are coupled with each other
by the magnetic field, and form three new dressed sates $|\varphi^{(\pm)\prime}\rangle$
and $|\varphi_D^\prime\rangle$. As a result,
although the probe beam only directly couples $|g\rangle_A|0\rangle_c$ to $|\varphi^{(\pm)}\rangle$, it can
induce three transitions from  $|g\rangle_A|0\rangle_c$ to
the three states $|\varphi^{(\pm)\prime}\rangle$ and $|\varphi_D^\prime\rangle$.}
\end{figure}

Two of the excited states of the total Hamiltonian $H$ are
\begin{equation}
|\varphi^{(\pm)}\rangle=\frac{1}{\sqrt{2}}(|g\rangle_{A}|1\rangle_{c}\pm|e_{x}\rangle_{A}|0\rangle_{c}),\label{phipm-1}
\end{equation}
with the corresponding eigenenergies
\begin{equation}
E_{\varphi^{(\pm)}}=\omega_{A}\pm\Omega/2.\label{epm-1}
\end{equation}
Here $|n\rangle_{c}$ ($n=0,1,2,...$) is the state of the cavity
field with $n$ photons. Without loss of generality, we have assumed that the Rabi frequency $\Omega$
is real. According to Eq.~(\ref{hp}), the probe beam couples the
ground state $|g\rangle_{A}|0\rangle_{c}$ to the states $|\varphi^{(\pm)}\rangle$.
Thus, our discussion in the above subsection yields that
the transmission spectrum has two peaks at
\begin{equation}
\omega_{p}=E_{\varphi^{(\pm)}}=\omega_{A}\pm\Omega/2,\label{omp-1}
\end{equation}
which is the so-called vacuum Rabi splitting [Fig.~\ref{fig:The-dressed-state-picture}(a)].

Furthermore, since $|\varphi^{\left(\pm\right)}\rangle$
are superpositions with equal weight of $|g\rangle_{A}|1\rangle_{c}$
and $|e_{x}\rangle_{A}|0\rangle_{c}$, the decay rates $\Gamma_{\varphi^{\left(\pm\right)}}$ of $|\varphi^{\left(\pm\right)}\rangle$
can be estimated as $(\gamma+\kappa)/2$,
with $\kappa$ and $\gamma$ being the  the leaking rate of the cavity photon and the
 spontaneous decay rate of atomic $^{3}{\rm P}_{1}$ state, respectively. Moreover, in current experiments we usually have $\kappa\gg\gamma$ \cite{fn1}. Therefore, the linewidth of each transmission peak is
\begin{equation}
\Gamma_{\varphi^{\left(+\right)}}=\Gamma_{\varphi^{\left(-\right)}}\approx\frac{\kappa}{2}.\label{gampm-1}
\end{equation}

For our system, the Raman beams $\alpha$ and $\beta$ can couple the state
$|e_{y}\rangle_{A}$ with the $^3{\rm S}_1$ state $|S\rangle_A$ and the long-lived $^{3}{\rm P}_{0}$ state
$|P\rangle_{A}$ [Fig.~\ref{fig:scheme1}(b)]. As a result, the total Hamiltonian $H$ defined in Eq.~(\ref{hh}) has another
eigenstate
\begin{equation}
|\varphi_{D}\rangle=|D\rangle_{A}|0\rangle_{c},\label{psiy-1}
\end{equation}
with
\begin{equation}
|D\rangle_{A}=\frac{g_{\beta}|e_{y}\rangle_{A}-g_{\alpha}|P\rangle_{A}}{\sqrt{|g_{\alpha}|^{2}+|g_{\beta}|^{2}}}\label{da-1}
\end{equation}
being the dark state corresponding to the $\Lambda$-type coupling
between states $|e_{y}\rangle_{A}$, $|S\rangle_A$ and $|P\rangle_{A}$. The eigenenergy of $H$ with respect to $|\varphi_{D}\rangle$ is
\begin{equation}
E_{\varphi_{D}}=\omega_{A}.\label{ey-1}
\end{equation}
However, since the probe Hamiltonian $H_{p}$ cannot induce the transition between $|g\rangle_{A}|0\rangle_{c}$ and $|\varphi_{D}\rangle$, there is no peak corresponding to the state $|\varphi_{D}\rangle$ in the transmission spectrum
[Fig.~\ref{fig:The-dressed-state-picture}(a)]. In this sense, $|\varphi_{D}\rangle$ is a dark state of our system.

\subsection{Case with $B\protect\neq0$: MIOT modulated by the Raman beams}

The MIOT effect occurs when a bias magnetic field $B$ along the $z$-direction
is applied. In this case, the magnetic field can induce Zeeman shifts
of the $^{3}{\rm P}_{1}$ states with magnetic quantum numbers $\pm1$,
and contributes a term $H_{B}=\Delta(|e_{1}\rangle\langle e_{1}|-|e_{-1}\rangle\langle e_{-1}|)/2$
in the Hamiltonian, where $\Delta=2\mu B$ is the Zeeman shift and
$\mu$ is the magnetic moment of $^{3}{\rm P}_{1}$ states. In
the basis $|e_{x,y,z}\rangle$, this term can be re-written as
\begin{equation}
H_{B}=-i\frac{\Delta}{2}|e_{x}\rangle_{A}\langle e_{y}|+\textrm{h.c.}.\label{hb}
\end{equation}
Accordingly, the self-Hamiltonian of our system becomes
\begin{equation}
H=H_{AP}+H_{{\rm Raman}}+H_{B},\label{hm}
\end{equation}
where $H_{AP}$ and $H_{{\rm Raman}}$ are given by Eqs.~(\ref{hap}) and (\ref{hra}), respectively.

When the magnetic field is so weak that $\Delta\ll\Omega$, we can
treat $H_{B}$ as a first-order perturbation and derive the approximated eigenstates
and eigenenergies of $H$. Since the Zeeman Hamiltonian $H_B$
couples the atomic states
$|e_{x}\rangle_{A}$ and $|e_{y}\rangle_{A}$, it can couple the dressed states $|\varphi^{(\pm)}\rangle$ with $|\varphi_{D}\rangle$. Therefore, the perturbative eigenstate of $H$ up to the first-order of $\Delta$ can be expressed as
\begin{eqnarray}
|\varphi_{D}^{\prime}\rangle & \approx & |\varphi_{D}\rangle+\frac{i\Delta g_{\beta}}{\sqrt{2}\Omega\sqrt{|g_{\alpha}|^{2}+|g_{\beta}|^{2}}}(|\varphi^{(+)}\rangle+|\varphi^{(-)}\rangle)\nonumber \\
 & = & |D\rangle_{A}|0\rangle_{c}+\frac{i\Delta g_{\beta}}{\Omega\sqrt{|g_{\alpha}|^{2}+|g_{\beta}|^{2}}}|g\rangle_{A}|1\rangle_{c},\label{psi0}
\end{eqnarray}
with corresponding eigenenergy
\begin{equation}
E_{\varphi_{D}^{\prime}}\approx\omega_{A}.\label{eyp}
\end{equation}
Eqs.~(\ref{psi0}) and (\ref{eyp}) show that, due to the magnetic field
induced coupling between $|\varphi^{(\pm)}\rangle$ and $|\varphi_{D}\rangle$,
the state $|g\rangle_{A}|1\rangle_{c}$ is mixed into $|\varphi_{D}\rangle$
with a small weight. Hence, the probe Hamiltonian
$H_{p}$ can induce the transition $|g\rangle_{A}|0\rangle_{c}\leftrightarrow|\varphi_{D}^{\prime}\rangle$ [Fig.~\ref{fig:The-dressed-state-picture}(b)] and result in a new peak in the transmission spectrum appearing at
\begin{equation}
\omega_{p}=E_{\varphi_{D}^{\prime}}\approx\omega_{A}.\label{np}
\end{equation}
The appearance of this transmission peak at the dark state energy is the MIOT effect.

The linewidth of this MIOT peak is just the decay rate
$\Gamma_{\varphi_{D}^{\prime}}$ of the state $|\varphi_{D}^{\prime}\rangle$.
According to Eq.~(\ref{psi0}) and Eq.~(\ref{da-1}), $|\varphi_{D}^{\prime}\rangle$ is the superposition of the long-lived state $|P\rangle_A|0\rangle_c$, as well as the states $|e_y\rangle_A|0\rangle_c$ and $|g\rangle_A|1\rangle_c$. The population of the latter two states are ${|g_{\beta}|^{2}}/({|g_{\alpha}|^{2}+|g_{\beta}|^{2}})$ and ${|g_{\beta}|^{2}}/({|g_{\alpha}|^{2}+|g_{\beta}|^{2}})(\Delta/\Omega)^2$, respectively. Thus, $\Gamma_{\varphi_{D}^{\prime}}$ can be estimated as
\begin{equation}
\Gamma_{\varphi_{D}^{\prime}}\approx\frac{|g_{\beta}|^{2}}{|g_{\alpha}|^{2}+|g_{\beta}|^{2}}\left[\gamma+\left(\frac{\Delta}{\Omega}\right)^{2}\kappa\right].\label{gpsip2}
\end{equation}
When $\Delta/\Omega$ is small enough, we further have
\begin{equation}
\Gamma_{\varphi_{D}^{\prime}}\approx\frac{|g_{\beta}|^{2}}{|g_{\alpha}|^{2}+|g_{\beta}|^{2}}\gamma.\label{gamapsi}
\end{equation}
Thus, the linewidth $\Gamma_{\varphi_{D}^{\prime}}$ of the MIOT
peak can be modulated by the Raman beams. In the absence of the Raman beam $\alpha$ (i.e., $g_\alpha=0$), the $^3{\rm P}_1$ state $|e_y\rangle_A$ is decoupled with the states $|S\rangle_A$ and $|P\rangle_A$, and Eq.~(\ref{gamapsi}) shows that in this case the linewidth $\Gamma_{\varphi_{D}^{\prime}}$ of the MIOT peak is just the spontaneous emission rate $\gamma$ of the $^{3}{\rm P}_{1}$ states.
Furthermore, when the Raman beam $\alpha$ is applied and the Rabi frequency $g_\alpha$ of this beam is much larger than  the beam $\beta$, i.e., $|g_\alpha|\gg|g_\beta|$, Eq.~(\ref{gamapsi}) indicates that the width $\Gamma_{\varphi_{D}^{\prime}}$ of the MIOT peak can be further decreased to a value much less than $\gamma$, i.e., we can have $\Gamma_{\varphi_{D}^{\prime}}\ll \gamma$. Therefore, with the help of the Raman beams $\alpha$ and $\beta$, one can significantly reduce the linewidth of the MIOT peak and realize an ultra-narrow spectrum. That is the basic principle of our scheme.

\section{ Ultra-narrow MIOT spectrum for $N$-atom system}

\subsection{Heisenberg-Langevin calculation}

The above single-atom analysis captures the most fundamental physics of our scheme for
MIOT with ultra-narrow linewidth, which also applies to the $N$-atom case. In this
section, we use the input-output theory to explicitly study the
transmission spectrum of the
$N$-atom system and investigate the detailed properties of  the
width, height,  location, and pulling coefficient for the MIOT-induced transmission peak.

We assume the cavity, Raman beams, and magnetic fields
are all uniformly coupled to the atoms. Thus the system Hamiltonian
$H$ defined in Eq.~(\ref{hm}) is generalized as
\begin{eqnarray}
H_{AP}^{(N)} & = & \omega_{A}\hat{a}_{x}^{\dagger}\hat{a}_{x}+\sum_{i=1}^{N}\omega_{A}(\vert e_{x}\rangle_{A}^{(i)}\langle e_{x}\vert+\vert e_{y}\rangle_{A}^{(i)}\langle e_{y}\vert)\nonumber \\
 &  & +\sum_{i=1}^{N}(\frac{\Omega}{2}\hat{a}_{x}\vert e_{x}\rangle_{A}^{(i)}\langle g_{i}\vert+\textrm{h.c.}),\label{Hn1}\\
H_{{\rm Raman}}^{(N)} & = & \sum_{i=1}^{N}\omega_{A}(\vert S\rangle_{A}^{(i)}\langle S\vert+\vert P\rangle_{A}^{(i)}\langle P\vert)\nonumber \\
 &  & +\sum_{i=1}^{N}(\frac{g_{\alpha}}{2}\vert S\rangle_{A}^{(i)}\langle e_{y}\vert+\frac{g_{\beta}}{2}\vert S\rangle_{A}^{(i)}\langle P\vert+\textrm{h.c.}),\nonumber \\
\label{Hn2}\\
H_{B}^{(N)} & = & -i\frac{\Delta}{2}\sum_{i=1}^{N}(\vert e_{x}\rangle_{A}^{(i)}\langle e_{y}\vert-\textrm{h.c.}),\label{Hn3}
\end{eqnarray}
where $|\rangle_A^{(i)}$ is the internal state of the {$i$}-th atom.
For the convenience of the following calculation,  we introduce the
the following collective atomic operators
\begin{align}
\hat{B}_{k} & =\frac{1}{\sqrt{N}}\sum_{i=1}^{N}\vert g\rangle_{A}^{(i)}\langle e_{k}\vert,\ (k=x,y),\label{B}\\
\hat{C}_{l} & =\frac{1}{\sqrt{N}}\sum_{i=1}^{N}\vert g\rangle_{A}^{(i)}\langle l\vert,\ (l=S,P).\label{C}
\end{align}
In addition, we assume the cavity field is weakly driven by the probe beam, so that
the atoms are mostly occupied by the ground states, and the populations
of the excited states can be neglected.
According to the Holstein-Primakoff approximation~\citep{Holstein_PR1940}, in this low-excitation regime ${\hat B}_{k=x,y}$ and ${\hat C}_{l=S,P}$ satisfy the bosonic commutation relation
\begin{equation}
[\hat{B}_{x(y)},\hat{B}^{\dagger}_{x(y)}]\approx1,\ [\hat{C}_{S(P)},\hat{C}_{S(P)}^{\dagger}]\approx1.
\end{equation}
In a rotated frame that the probe Hamiltonian is time-independent
(see Appendix~\ref{appnB}), the Heisenberg-Langevin
equations  for  our system are derived as
\begin{eqnarray}
i\dot{a}_{x} & = & (\delta_{p}-i\frac{\kappa}{2})a_{x}+\frac{\Omega_{N}}{2}B_{x}+i\sqrt{\frac{\kappa I_{p}}{2\omega_{p}}},\label{HL1}\\
i\dot{B}_{x} & = & (\delta_{p}-i\frac{\gamma}{2})B_{x}+\frac{\Omega_{N}}{2}a_{x}-i\frac{\Delta}{2}B_{y},\label{HL2}\\
i\dot{B}_{y} & = & (\delta_{p}-i\frac{\gamma}{2})B_{y}+\frac{g_{\alpha}^{*}}{2}C_{S}+i\frac{\Delta}{2}B_{x},\label{HL3}\\
i\dot{C}_{S} & = & (\delta_{p}-i\frac{\Gamma}{2})C_{S}+\frac{g_{\alpha}}{2}B_{y}+\frac{g_{\beta}}{2}C_{P},\label{HL4}\\
i\dot{C}_{P} & = & \delta_{p}C_{P}+\frac{g_{\beta}^{\ast}}{2}C_{S}.\label{HL5}
\end{eqnarray}
Here, we denote $O\equiv\langle\hat{O}\rangle$ as the expectation
value of the operator $\hat{O}$, and $\dot{O}$ is the time derivative of $O$.
$\Gamma$ is the decay rate of atomic state $\vert S\rangle_{A}$,
while $\kappa$ and $\gamma$ are the decay rates of the cavity mode and the atomic $^{3}{\rm P}_{1}$ state, respectively, as defined above.
In addition, the cavity-probe detuning $\delta_{p}$ and collective Rabi frequency $\Omega_{N}$ are defined as
\begin{equation}
\delta_{p}\equiv\omega_{A}-\omega_{p}\label{deltap}
\end{equation}
and
\begin{equation}
\Omega_{N}\equiv\sqrt{N}\Omega,
\end{equation}
respectively. In deriving Eqs.~(\ref{HL1}-\ref{HL5}),
we assume the number of the atoms in the excited states is quite small,
such that the ground state population $\langle\sum_{i=1}^{N}\vert g\rangle_{A}^{(i)}\langle g\vert\rangle\approx N$,
and all the transitions terms between excited states (e.g., $\sum_{i=1}^{N}\vert e_{x}\rangle_{A}^{(i)}\langle S\vert/\sqrt{N}$
and $\sum_{i=1}^{N}\vert e_{x}\rangle_{A}^{(i)}\langle P\vert/\sqrt{N}$) are negligible.

%In this work, we choose $\kappa_m=\kappa/2$ and the decay rate of state $\vert S\rangle_{A}$ as $\Gamma=2\pi\times5.3$MHz \citep{Vanier_2016,Courtillot_EPJD2005}.

For our system, the transmission spectrum of the probe beam is proportional to the ratio $P_{T}$ between the
steady-state transmitted power and the incident probe beam
power~\citep{Collett_PRA198485,Walls_2008}, which is a function of the probe frequency $\omega_{p}$ and can be expressed as
\begin{equation}
P_{T}=\frac{\kappa\omega_{p}}{2I_{p}}\vert a^{\rm (st)}_{x}\vert^{2}.\label{PT1}
\end{equation}
Here $a^{\rm (st)}_{x}$ is the value of $a_x$ for the steady-state solution of Eqs.~(\ref{HL1}-\ref{HL5}). With direct calculations, we find that
$P_{T}$ can be analytically expressed as
\begin{align}
P_{T} & =\frac{\kappa^{2}}{4}\left|\frac{(\delta_{p}-i\frac{\gamma}{2})F-\frac{\Delta^{2}}{4}R}{(\delta_{p}-i\frac{\kappa}{2})[(\delta_{p}-i\frac{\gamma}{2})F-\frac{\Delta^{2}}{4}R]-\frac{\Omega_{N}^{2}}{4}F}\right|^{2},\label{PT2}
\end{align}
where $R$ and $F$ are given by
\begin{eqnarray}
R & = & \eta-\frac{4\delta_{p}(\delta_{p}-i\frac{\Gamma}{2})}{\vert g_{\alpha}\vert^{2}+\vert g_{\beta}\vert^{2}},\label{R}\\
F & = & \delta_{p}\left(R+1-\eta\right)-i\frac{\gamma}{2}R\label{F}
\end{eqnarray}
with the factor $\eta$ being defined as
\begin{equation}
\eta=\frac{|g_{\beta}|^{2}}{|g_{\alpha}|^{2}+|g_{\beta}|^{2}}.\label{etaeta}
\end{equation}

In this work, we mostly concern the transmission peak induced by the MIOT effect, which appears at $\delta_{p}\approx 0$. Specifically, we focus on the following three properties of the MIOT peak:
\begin{itemize}
\item[{\bf (i)}] {\bf MIOT peak height  $P^{\rm (max)}_T$:}  the maximum  value of $P_T$ in the region of the transmission peak around $\delta_p\approx 0$.

\item[{\bf (ii)}] {\bf MIOT linewidth $W_{\rm MIOT}$:}  the full width of this transmission spectrum at its half maximum.

\item[{\bf (iii)}] {\bf MIOT peak position $\delta_p^{\rm MIOT}$:} the value of $\delta_p$ where we have $P_T=P^{\rm (max)}_T$.
 \end{itemize}

The exact values of these parameters can be derived directly with Eq.~(\ref{PT2}).
Moreover, we find that under the condition $\Omega_{N}\gg\Delta\gg\{\gamma,\kappa\}$, the exact expression of Eq.~(\ref{PT2}) can be further approximated as
\begin{equation}
P_{T}\lvert_{\delta_p\approx0}\approx\frac{\kappa^{2}}{4}\left|\frac{\eta\bar{\Delta}^{2}/\Omega_{N}^{2}}{\delta_{p}-\frac{i}{2}\eta(\gamma+\kappa\bar{\Delta}^{2}/\Omega_{N}^{2})}\right|^{2}\label{PT3}
\end{equation}
with
\begin{eqnarray}
\bar{\Delta}\equiv\sqrt{\Delta^{2}+\gamma^{2}}.
\end{eqnarray}
Using Eq.~(\ref{PT3}) we can further derive the approximated expressions for the position, linewidth and height of the MIOT peak:
 \begin{align}
 \delta_p^{\rm MIOT}&\approx 0,\label{ddapp}\\
W_{\rm MIOT}& \approx \eta \left(\gamma+\kappa\frac{\bar{\Delta}^{2}}{\Omega_{N}^{2}}\right),\label{width}\\
P^{\rm (max)}_T&  \approx \left|\frac{\bar{\Delta}^{2}}{\bar{\Delta}^{2}+\frac{\gamma}{\kappa}\Omega_{N}^{2}}\right|^{2}.\label{height}
\end{align}

In the following three subsections, we will investigate the properties of the linewidth $W_{P_T}$, height $P^{\rm (max)}_T$, and the pulling coefficient which describes the variation of the MIOT peak position $\delta_p^{\rm MIOT}$ with respect to the frequency fluctuations of the cavity-mode and the Raman beams.

\subsection{Linewidth of the MIOT peak}

Our result in Eq.~(\ref{width}) yields that the linewidth $W_{\rm MIOT}$ of the MIOT peak can be modulated by the Raman beams, and reduced by a factor $\eta$ from the cases without the Raman beams. In particular, when $|g_\alpha|\gg |g_\beta|$ (i.e., $\eta\ll 1$), this reduction effect is very significant.
All these conclusions are consistent with our analysis of the single-atom case in Sec.~II.

\begin{figure}[tbp]
\begin{center}
\includegraphics[width=6.5cm]{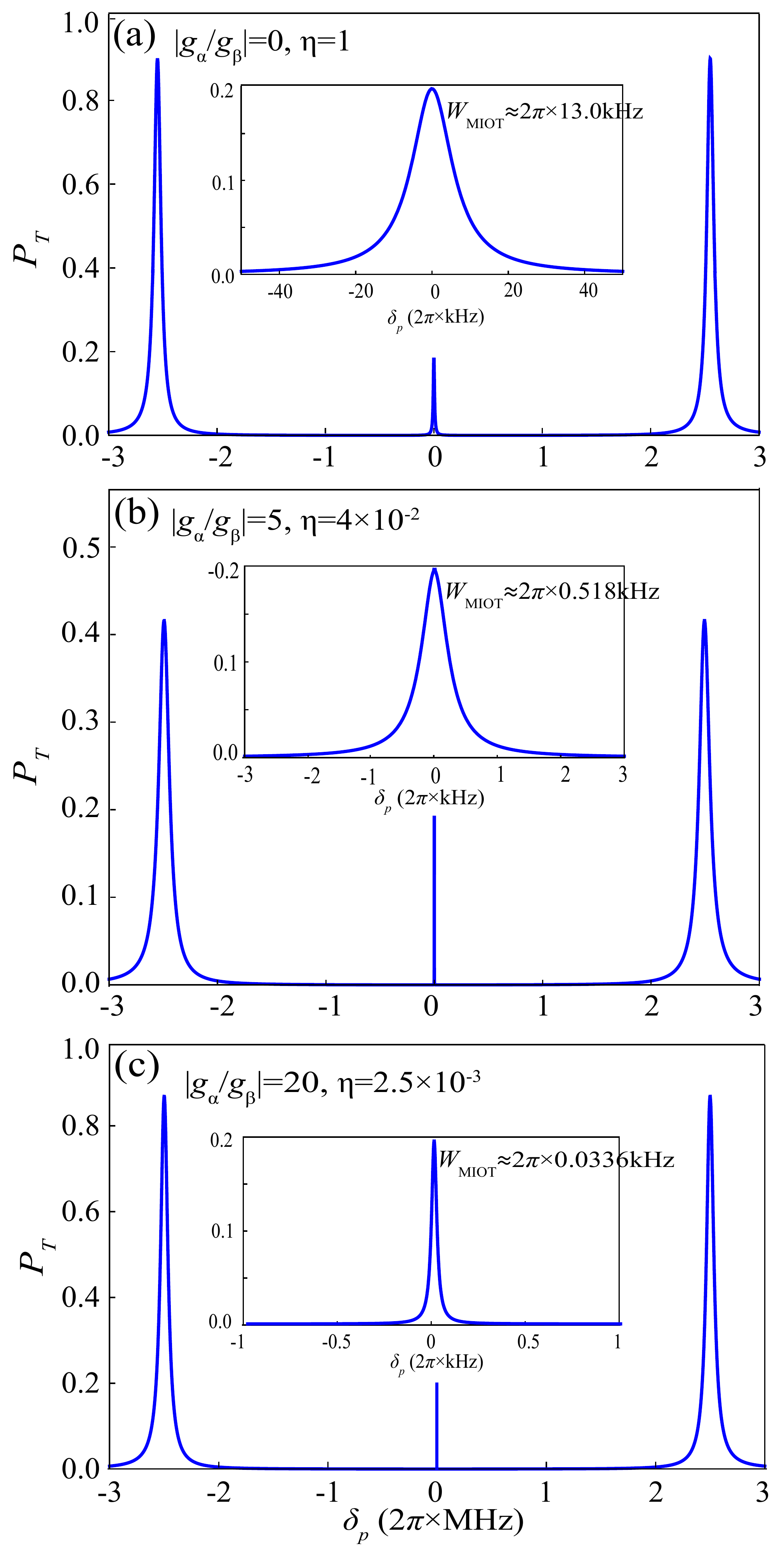}
\end{center}
\caption{\label{fig:PT} (Color online)
The relative transmission power $P_{T}$ of the MIOT peak as a function of the detuning
$\delta_p$, which is given by the exact result Eq.~(\ref{PT2}).
{\bf (a):} The result with no Raman beams ($g_\alpha=g_\beta=0$).
{\bf (b):} The result for $g_{\alpha}=2\pi\times10$MHz and $g_{\beta}=2\pi\times2$MHz ($|g_\alpha/g_\beta|=5$ or $\eta\approx 4\times 10^{-2}$).
{\bf (c):} The results for $g_{\alpha}=2\pi\times40$MHz and $g_{\beta}=2\pi\times2$MHz ($|g_\alpha/g_\beta|=20$ or $\eta\approx 2.5\times 10^{-3}$).
In our calculation we
consider the system with $^{88}$Sr atoms, and
take $\Omega_{N}=2\pi\times5$MHz, $\Delta=2\pi\times1$MHz, $\kappa=2\pi\times150$kHz and $\gamma=2\pi\times7.5$kHz as in
the experimental work~\citep{Winchester_PRL2017}
of $^{88}$Sr atoms.
The decay rate $\Gamma$
 of the $^3{\rm S}_1$ state is taken as   $\Gamma=2\pi\times5.3$MHz
\citep{Vanier_2016,Courtillot_EPJD2005}.
The insets in each figure are the zoom-in views of the MIOT peak with the exact values of $W_{\rm MIOT}$ are also indicated. In addition, the values of $W_{\rm MIOT}$ given by the approximated expression Eq.~(\ref{width}) for the cases of (a), (b) and (c) are  $W_{\rm MIOT}\approx 2\pi\times13.5$kHz, $2\pi\times0.519$kHz and $2\pi\times0.0337$kHz, respectively.}
\end{figure}

To illustrate the Raman-beam-induced reduction of the MIOT linewidth, in Fig.~\ref{fig:PT} we show the relative transmission power $P_{T}$ obtained from the exact expression Eq.~(\ref{PT2}) with different Rabi frequencies of the Raman beams. We can see there are three transmission peaks in each figure. The narrow peak around $\delta_p=0$ is induced by the MIOT effect, and the other two peaks correspond to the vacuum Rabi splitting, as discussed in Sec.~II \cite{peak}. In Fig.~\ref{fig:PT} we also give the exact MIOT linewidth $W_{\rm MIOT}$ for each case.

\begin{figure}[tbp]
\begin{centering}
\includegraphics[width=6.5cm]{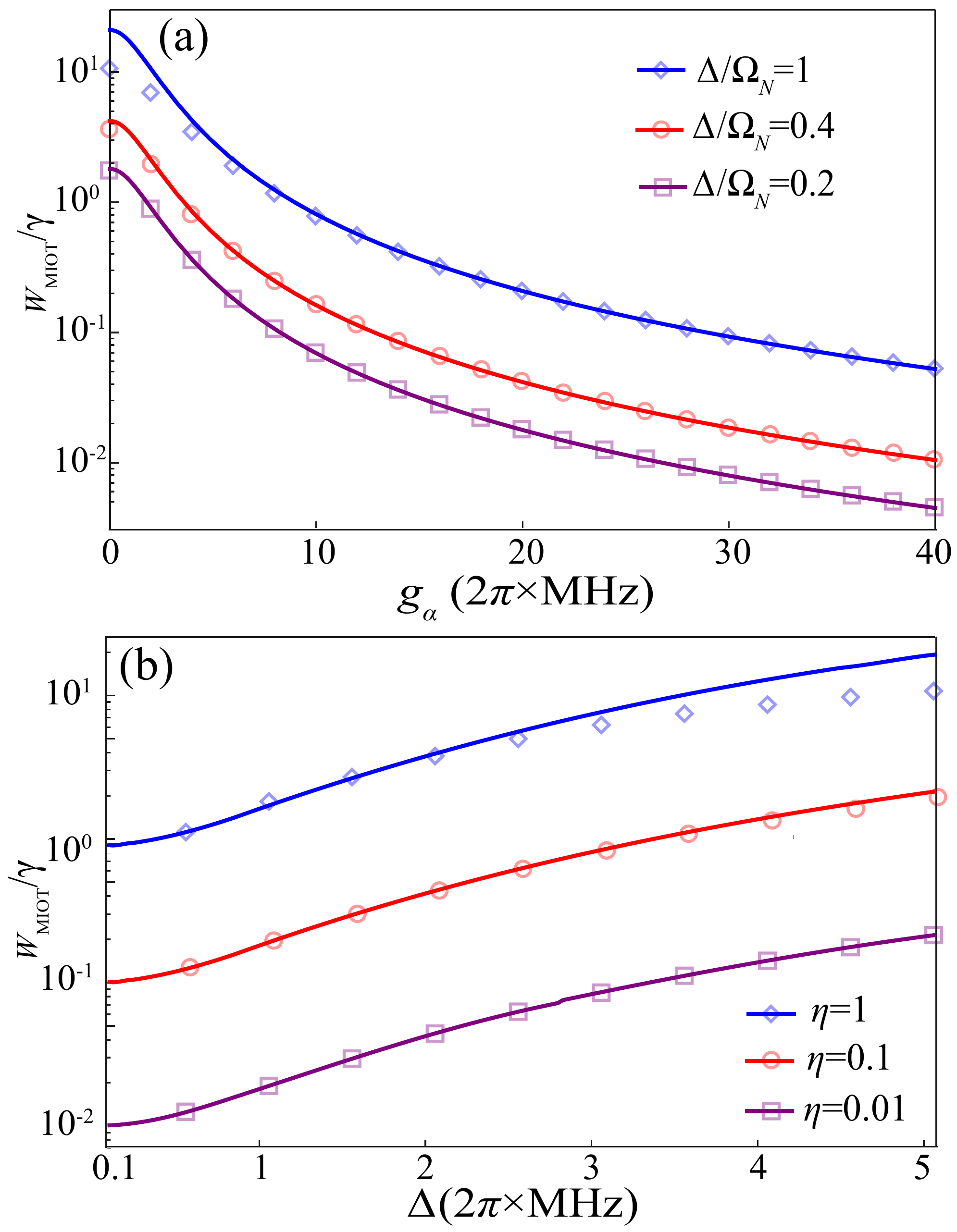}
\par\end{centering}
\caption{\label{fig:kD} (Color online)
{\bf (a):} The MIOT linewidth $W_{\rm MIOT}$ as a function of the Rabi
frequency $g_{\alpha}$ of the Raman beam  with various Zeeman energy $\Delta$.
{\bf (b):} $W_{\rm MIOT}$ as a function of the Zeeman energy $\Delta$  with various Rabi
frequency $g_{\alpha}$. The other parameters are same as in Fig.~\ref{fig:PT}.
Here we show both the exact value of $W_{\rm MIOT}$ given by
the exact solution of the  Heisenberg-Langevin equations (\ref{HL1}-\ref{HL5})
(open diamonds,  circles, and  squares) and the results from the approximation Eq.~(\ref{width}) (the correspongding solid lines).}
\end{figure}

\begin{figure}[tbp]
\begin{center}
\includegraphics[width=6.5cm]{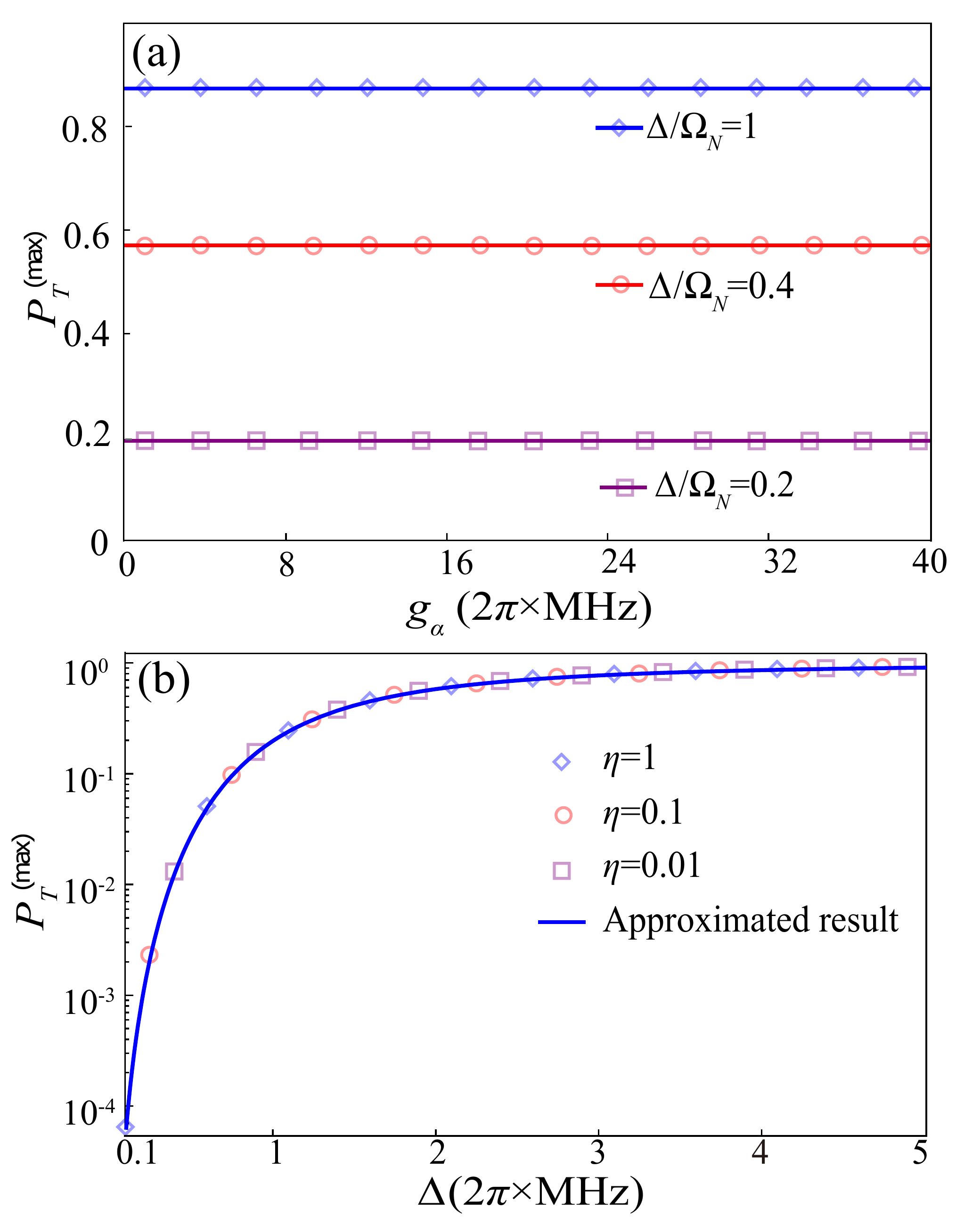}
\end{center}
\caption{\label{fig:Power} (Color online) The height $P^{\rm (max)}_T$ of the MIOT peak given by the exact calculation (open diamonds, circles, and squares) and the approximated expression Eq.~(\ref{height}) (the corresponding solid lines). {\bf (a):} $P^{\rm (max)}_T$ as a function of the the Rabi frequency $g_{\alpha}$, with $\Delta/\Omega_{N}=$1, 0.4 and 0.2. {\bf (b):} $P^{\rm (max)}_T$ as a function of the Zeeman splitting $\Delta$. Here we show the exact results for $\eta=$1, 0.1 and 0.01. Notice that the approximated result is independent of $\eta$. Other parameters are the same as Fig.~(\ref{fig:kD}).}
\end{figure}

In Fig.~\ref{fig:PT}(a) we show the results for the case without Raman beams ($g_{\alpha}=g_{\beta}=0$), i.e., the case of the experiment in Ref.~\citep{Winchester_PRL2017}. It is shown that
the MIOT linewidth $W_{\rm MIOT}$ given by
our calculation is  about 1.7 times of the decay rate $\gamma$ of the $^3{\rm P}_1$ state, which is consistent with Ref.~\citep{Winchester_PRL2017} and our approximate result in Eq.~(\ref{width}). In Fig.~\ref{fig:PT}(b, c) we illustrate the transmission spectrum for typical cases where the Raman beams are applied, with the parameter $\eta$ defined in Eq.~(\ref{etaeta}) being about $4\times 10^{-2}$ and $2.5\times 10^{-3}$, respectively. It is clearly shown that by decreasing the value of $\eta$ or increasing the ratio $|g_\alpha/g_\beta|$, one can significantly decrease the linewidth $W_{\rm MIOT}$. For instance, when $\eta\approx2.5\times 10^{-3}$ ($|g_\alpha/g_\beta|=20$), $W_{\rm MIOT}$ is decreased to about $(2\pi)3.4\times 10^{-2}{\rm kHz}$ [Fig.~\ref{fig:PT}(c)].

We further illustrate the MIOT linewidth $W_{\rm MIOT}$ as functions of the Rabi frequency $g_\alpha$ [Fig.~\ref{fig:kD}(a)] and the Zeeman shift $\Delta$ [Fig.~\ref{fig:kD}(b)], and compare the exact values of $W_{\rm MIOT}$ with the approximate results given by Eq.~(\ref{width}). It is clearly shown that this approximation works quite well. In addition, according to Fig.~\ref{fig:kD}, one can decrease the MIOT linewidth $W_{\rm MIOT}$ to be significantly below the nature linewidth $\gamma$ of the $^{3}{\rm P}_{1}$ state, by either increasing the ratio $g_\alpha/g_\beta$ (i.e., decreasing the parameter $\eta$) or decreasing the Zeeman shift $\Delta$ (i.e., decreasing the magnetic field $B$).
Nevertheless, in realistic cases one cannot unlimitedly decrease $\Delta$, because of the relative output power fast declines with $\Delta$
(as we will discuss below). Therefore, the effective approach of the reduction of the MIOT linewidth is increasing the ratio $g_{\alpha}/g_{\beta}$ or the parameter $\eta$.

\subsection{Height of the MIOT peak}

\begin{figure*}[tbp]
\begin{centering}
\includegraphics[scale=0.3]{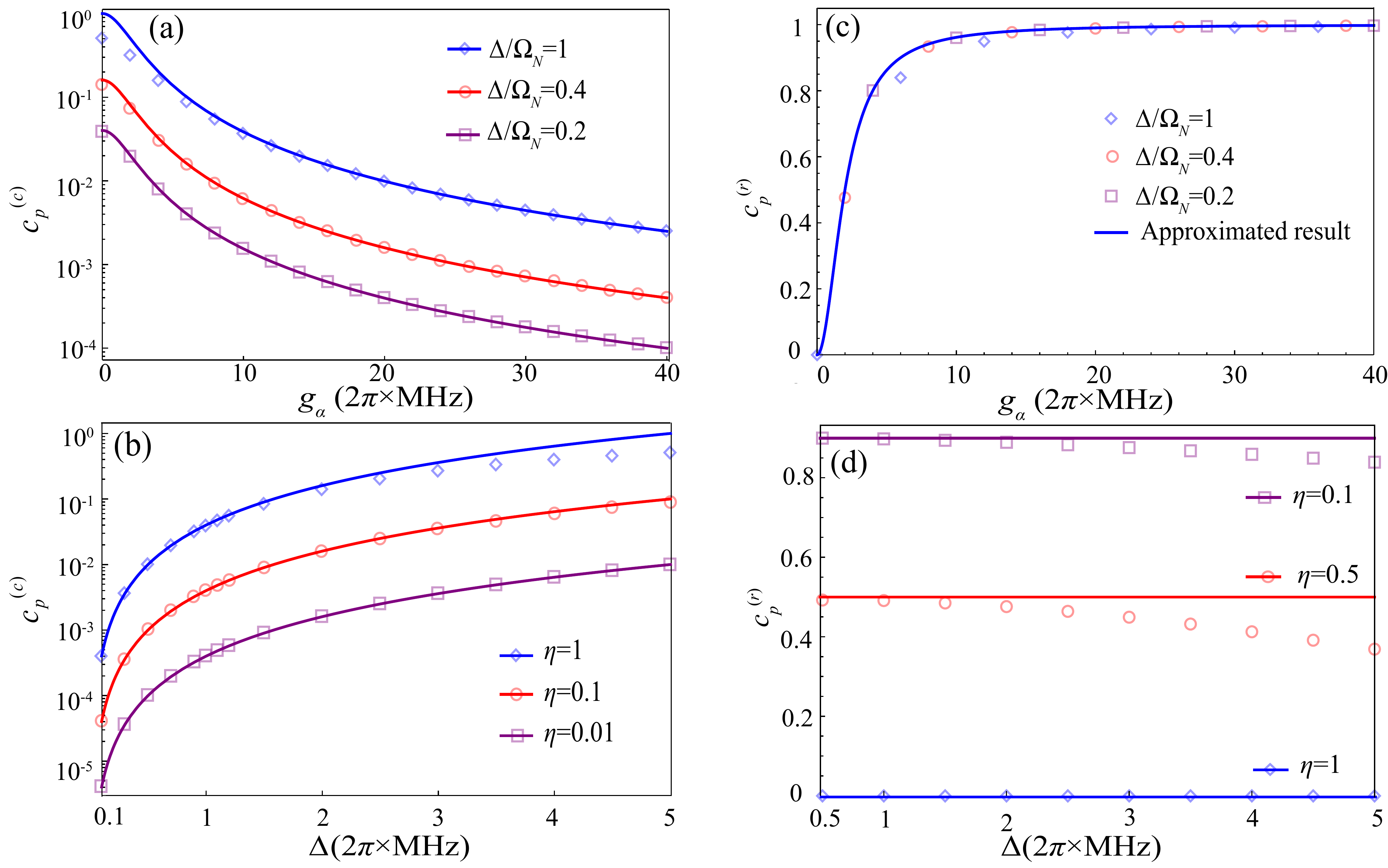}
\par\end{centering}
\caption{\label{fig:pulling} (Color online) The pulling coefficients $c_{p}^{ (c)}$ and $c_{p}^{ (r)}$ given by the exact calculation (open diamonds, circles,  and squares) and the approximated expression (the corresponding solid lines) from Eq.~(\ref{eq:cp}) and Eq.~(\ref{eq:pulling_alpha}), respectively. {\bf (a, c):} $c_{p}^{ (c)}$ and $c_{p}^{ (r)}$ as functions of the the Rabi frequency $g_{\alpha}$ with $\Delta/\Omega_{N}=1, 0.4, 0.2$. {\bf (b, d):} $c_{p}^{ (c)}$ and $c_{p}^{ (r)}$ as functions of the Zeeman splitting $\Delta$ with $\eta=1, 0.1, 0.01$ and $\eta=1, 0.5, 0.1$, respectively. Here we take $\Omega_{N}=2\pi\times5$MHz and $g_{\beta}=2\pi\times2$MHz. The other parameters are the same as Fig.~\ref{fig:Power}.}
\end{figure*}

Now we consider the height $P^{\rm (max)}_T$ of the MIOT peak. In Fig.~\ref{fig:Power} we compare the exact value of $P^{\rm (max)}_T$ and the approximate results given by Eq.~(\ref{height}) for various cases, and it is shown that this approximation works very well.
Furthermore, Eq.~(\ref{height}) and Fig.~\ref{fig:Power}(a) show that the MIOT peak height is independent of the Rabi frequencies $g_{\alpha,\beta}$. This  is an advantage of our approach, because it yields that when the width of the MIOT peak is decreased by the Raman beams, the height of this peak is not suppressed.

The fact that $P^{\rm (max)}_T$ is independent of $g_{\alpha,\beta}$ can be understood with the following analysis for the single-atom system of Sec.~II. Since the MIOT peak is the transmission peak corresponding to the probe-beam-induced transition between the ground state $|g\rangle_A|0\rangle_c$ and the dark state $|\varphi_{D}^{\prime}\rangle$ of Eq.~(\ref{psi0}), the height of this peak can be estimated as proportional to $|\langle \varphi_{D}^{\prime}|H_p|g\rangle_A|0\rangle_c|^2/\Gamma_{\varphi_{D}^{\prime}}$, with $H_p$ the probe beam Hamiltonian defined in Eq.~(\ref{hp}) and $\Gamma_{\varphi_D^\prime}$ the dark state decay rate given in Eq.~(\ref{gpsip2}). Using Eqs.~(\ref{hp}, \ref{psi0}, \ref{gpsip2}) we can find that both $|\langle \varphi_{D}^{\prime}|H_p|g\rangle_A|0\rangle_c|^2$ and $\Gamma_{\varphi_{D}^{\prime}}$ depends on the Raman-beam Rabi frequencies $g_{\alpha,\beta}$ via a coefficient $\eta$, thus the ratio $|\langle \varphi_{D}^{\prime}|H_p|g\rangle_A|0\rangle_c|^2/\Gamma_{\varphi_{D}^{\prime}}$ is independent of $g_{\alpha,\beta}$.

In addition, as shown in Eq.~(\ref{height}) and Fig.~\ref{fig:Power}(b), height of the MIOT peak $P^{\rm (max)}_T$ is an increasing function of the Zeeman energy gap $\Delta$. Explicitly, $P^{\rm (max)}_T$ takes a very small value $1/(1+\Omega_{N}^{2}/\kappa\gamma)$ in the limit $\Delta\rightarrow 0$ and approaches its maximum value 1 only when $\Delta\gg \Omega_{N}\sqrt{\gamma/\kappa}$. Therefore, as mentioned in Sec.~III.B, in order to ensure the MIOT peak is high enough in realistic cases, the Zeeman energy gap $\Delta$ should be large enough.

\subsection{Position of the MIOT peak}

As shown in Sec.~III.A, the exact position of the MIOT peak $\delta_p^{\rm MIOT}$ depends on the parameters $(\Delta,\Omega_N,\kappa,\gamma,g_{\alpha,\beta})$, and can be calculated via the steady-state solution Eq.~(\ref{PT2}) of the Heisenberg-Langevin equations. Moreover, we have $\delta_p^{\rm MIOT}\approx 0$ under the condition $\Omega_{N}\gg\Delta\gg\{\gamma,\kappa\}$, as shown in Eq.~(\ref{ddapp}).

In this section, we focus on the stability of the MIOT peak position. In
the above discussions we have assumed that the cavity mode is exactly resonant with the atomic $^1{\rm S}_0\leftrightarrow{^3{\rm P}_1}$ transition, and the Raman beams $\alpha$ and $\beta$ are exactly resonant with the $^3{\rm P}_1\leftrightarrow{^3{\rm S}_1}$ and $^3{\rm S}_1\leftrightarrow{^3{\rm P}_0}$ transitions, respectively. However, in realistic cases both the cavity-mode and the Raman-beams have random fluctuations. Namely, the exact cavity frequency $\omega_c$ and the Raman-beam frequencies $\omega_{\alpha,\beta}$ are given by
\begin{eqnarray}
\omega_c&=&\omega_A+\Delta_c,\\
\omega_\alpha&=&\omega_{SA}+\frac{\Delta_{t}}{2}+\frac{\Delta_{r}}{2},\\
\omega_\beta&=&\omega_{SP}+\frac{\Delta_{t}}{2}-\frac{\Delta_{r}}{2},
\end{eqnarray}
where
$\omega_{SA(SP)}$ is the frequency of the atomic $^{3}\mathrm{P}_{1}\leftrightarrow{^{3}\mathrm{S}_{1}}$ ($^{3}\mathrm{S}_{1}\leftrightarrow{^{3}\mathrm{P}_{0}})$ transitions, respectively, and $\Delta_{c,t,r}$ are
small stochastic fluctuations. Notice that $\Delta_t=\omega_\alpha+\omega_\beta-(\omega_{SA}+\omega_{SP})$ is actually the the fluctuation of the {\it total frequency} of the two Raman beams (i.e., the random one-photon detuning, and $\Delta_{r}=\omega_\alpha-\omega_\beta-(\omega_{SA}-\omega_{SP})$ is the fluctuation of the {\it frequency difference} of the two Raman beams (i.e., the random two-photon detuning).

The frequency fluctuations $\Delta_{c,t,r}$ can cause an unknown shift of the position $\delta_p^{\rm MIOT}$ of the MIOT peak, and thus $\delta_p^{\rm MIOT}$ should be a function of these fluctuations, i.e., we have $\delta_p^{\rm MIOT}=\delta_p^{\rm MIOT}(\Delta_c,\Delta_t,\Delta_r)$.
This random shift of the MIOT peak position can be described by the pulling coefficients $c_{p}^{ (j)}$
($j=c,t,r$) which are defined as
\begin{eqnarray}
&&c_{p}^{ (j)}=\Bigg|\left.\frac{\partial}{\partial \Delta_j}\delta_p^{\rm MIOT}(\Delta_c,\Delta_t,\Delta_r)\right|_{\Delta_c=\Delta_t=\Delta_r=0}\Bigg|,\nonumber\\
&&(j=c,t,r).
\end{eqnarray}
If $c_{p}^{(j)}$ ($j=c,t,r$) is small, it means the system is robust against the frequency fluctuation $\Delta_j$.

We can numerically derive $\delta_p^{\rm MIOT}(\Delta_c,\Delta_t,\Delta_r)$ and the pulling coefficients
$c_{p}^{(c,t,r)}$  by replacing the terms $\omega_Aa_x^\dagger a_x$ in Eq.~(\ref{Hn1}) and $\omega_{A}(\vert S\rangle_{A}^{(i)}\langle S\vert+\vert P\rangle_{A}^{(i)}\langle P\vert)$ in Eq.~(\ref{Hn2}) with
$\omega_ca_x^\dagger a_x$ and $[\omega_A-(\Delta_t+\Delta_r)/2]\vert S\rangle_{A}^{(i)}\langle S\vert+(\omega_A-\Delta_r)\vert P\rangle_{A}^{(i)}\langle P\vert$, respectively, and solving the corresponding Heisenberg-Langevin equation. Besides, similar as Sec.~III.A,
we can derive approximate expressions of the pulling coefficients for the cases with $\Omega_{N}\gg\Delta\gg\{\gamma,\kappa\}$:
\begin{eqnarray}
&&c_{p}^{(c)}\approx\left(\frac{\bar{\Delta}}{\Omega_{N}}\right)^2\eta,\label{eq:cp}\\
&&c_{p}^{\left(t\right)}\approx 0,\label{cpalpha}\\
&&c_{p}^{\left(r\right)}\approx 1-\eta.\label{eq:pulling_alpha}
\end{eqnarray}
In Fig.~\ref{fig:pulling} we show the pulling coefficients, $c_{p}^{(c)}$ and $c_{p}^{(r)}$, as functions of
the Rabi frequency $g_\alpha$ of Raman beams and the Zeeman splitting energy $\Delta$. It is shown that the exact values
agrees very well with the approximate results in Eq.~(\ref{eq:cp}) and Eq.~(\ref{eq:pulling_alpha}). In addition, our numerical calculation shows that exact value of $c_{p}^{(t)}$ is of the order of $10^{-14}$, which is consistent with the approximated result Eq.~(\ref{cpalpha}).

Furthermore, as shown Eq.~(\ref{width}), the ultra-narrow MIOT linewidth
appears when $\eta\ll 1$ and $|\bar{\Delta}|/|\Omega_{N}|\ll 1$. Eqs.~(\ref{eq:cp}-\ref{eq:pulling_alpha}) yield in this parameter region, and we have $c_{p}^{(c)}\ll 1$,  $c_{p}^{(t)}\ll 1$ and $c_{p}^{(r)}\approx 1$.
Thus, the ultra-narrow MIOT transmission spectrum is
very stable to the fluctuations of the cavity frequency and the
one-photon detuning of the Raman beams.
On the other hand, the fluctuation  $\Delta_{r}$ of the frequency difference of the two Raman beams
can lead to an uncertainty of the MIOT peak position, which is almost same as $|\Delta_{r}|$. In current experiments one can suppress $|\Delta_{r}|$ to the Hz (or even lower)  level via various techniques, e.g., locking the two Raman beams with two comb lines of an optical frequency comb or two modes of the same cavity.
Therefore, the above MIOT-peak position uncertainty can also be of this order.

\section{Conclusions  \label{sec:Conclusions-and-discussions}}

In this work, we show that an MIOT effect with ultra-narrow spectrum can be realized
in the cavity QED system with cold alkaline-earth (like) atoms dressed by two Raman beams. In our scheme
the linewidth of the MIOT-induced transmission peak can be reduced to the Hz or even lower level, which
is at least three orders smaller than that in the current experiment of MIOT
\citep{Winchester_PRL2017}.  Meanwhile, the heigh of this transmission peak is almost unchanged by the Raman beams, and the fluctuation of the peak position can be same as the one of the frequency difference of the two Raman beams.
Our scheme may be helpful for studies of precision measurement and other quantum optical processes based on cavity QED, e.g., the superradiant lasing \citep{Meiser_PRL2009,Chen_CSB2009,Bohnet_Nature2012,Norcia_PRX2016}.

\section{Acknowledgements}
We thank Prof.~Florian Schreck  for very helpful discussions.
G. Dong is supported by NSFC (Grant No.~11534002), NSAF (Grant No.~U1730449 \& No.~U1530401), and
National Basic Research Program of China (Grant No.~2016YFA0301201).
D. Xu is supported by NSFC (Grant No.~11705008).
P. Zhang is supported by the National Key R$\&$D Program of China (Grant No. 2018YFA0306502), NSFC Grant No.  11674393, and NSAF Grant No. U1930201.

\appendix

\section{Selection rule of the Raman beam $\alpha$ \label{appnA}}

In this appendix we derive the selection rule for the Raman beam $\alpha$, which couple the atomic $^3{\rm P}_1$ states to the $^{3}{\rm S}_{1}$ states. Since this beam is polarized along the $x$-direction, the Hamiltonian $H_\alpha$ for the coupling between this beam and the atom is proportional to
${\hat {\bf d}}\cdot{\bf e}_x$, where ${\hat {\bf d}}$ and ${\bf e}_x$ are the total electric dipole operator  and the unit vector along the $x$-direction, respectively. Introducing the complex unit vectors ${\bf e}_{\pm}=({\bf e}_x\pm i {\bf e}_y)/\sqrt{2}$, we find that
\begin{eqnarray}
H_\alpha\propto \left({\hat {\bf d}}\cdot{\bf e}_++{\hat {\bf d}}\cdot{\bf e}_-\right).\label{aa2}
\end{eqnarray}
The physical meaning of the above result is that the beam polarized along the $x$-direction is actually the combination of two beams with $\sigma_+$- and  $\sigma_-$-polarization.

Moreover, the calculation based on the angular momentum theory yields
\begin{eqnarray}
_A\langle s_0 |{\hat {\bf d}}\cdot{\bf e}_+|e_{-1}\rangle_A&=&_A\langle s_0 |{\hat {\bf d}}\cdot{\bf e}_-|e_{+1}\rangle_A\neq 0,\label{a1}
\end{eqnarray}
and
\begin{eqnarray}
_A\langle s_j |{\hat {\bf d}}\cdot{\bf e}_+|e_{\pm 1}\rangle_A&=&_A\langle s_j |{\hat {\bf d}}\cdot{\bf e}_-|e_{\pm 1}\rangle_A= 0,  (j=\pm 1),\nonumber\\
\label{a2}
\end{eqnarray}
where $|e_{j}\rangle_A$ and $|s_j\rangle_A$ ($j=0,\pm 1$) are the $^3{\rm P}_1$ and $^3{\rm S}_1$ states with magnetic quantum number $m_J=j$, respectively.
%In Fig.~\ref{cg} we further show the Clebsch-Gordan (CG) coefficients we calculated, which are related to the process in this work and proportional to the corresponding matrix elements of the total electric dipole operator.

Using Eq. (\ref{aa2}) and Eqs. (\ref{a1}, \ref{a2}), we can obtain
\begin{eqnarray}
_A\langle s_0 |H_\alpha|e_y\rangle_A\neq 0,\ \
_A\langle s_\pm |H_\alpha|e_y\rangle_A= 0, \label{aa3}
\end{eqnarray}
and
\begin{eqnarray}
_A\langle s_j |H_\alpha|e_x\rangle_A= 0,\ \ (j=0,\pm 1), \label{aa4}
\end{eqnarray}
with the $^3{\rm P}_1$ states $|e_{x,y}\rangle_A$ being defined in Eqs.~(\ref{exa}, \ref{eya}) in Sec.~II. Eqs.~(\ref{aa3}, \ref{aa4}) clearly show that the Raman beam $\alpha$ couples $|e_{y}\rangle_{A}$
to the $^{3}{\rm S}_{1}$ state $|s_0\rangle_{A}$ with magnetic quantum number $m_{J}=0$,
(i.e., the state $|S\rangle_{A}$ in Sec.~II), and does not couple $|e_{x}\rangle_{A}$
to any $^{3}{\rm S}_{1}$ state.

\section{Rotated frame for the Heisenberg-Langevin equations \label{appnB}}

The Heisenberg-Langevin
equations (\ref{HL1}-\ref{HL5}) of Sec.~III are derived in the rotated frame
where the  state $|\Psi(t)\rangle_I$ at time $t$ satisfies
\begin{eqnarray}
|\Psi (t)\rangle_I=e^{iH_{0}^{\left(N\right)}t} |\Psi (t)\rangle_S,
\end{eqnarray}
with $|\Psi (t)\rangle_S$ being the state in the Schr${\ddot{\rm o}}$dinger picture, and the Hamiltonian $H_{0}^{\left(N\right)}$ being defined as
(after the Holstein-Primakoff approximation)
\begin{eqnarray}
H_{0}^{\left(N\right)} & = & \omega_{p}\left(\hat{a}_{x}^{\dagger}\hat{a}_{x}+\hat{B}_{x}^{\dagger}\hat{B}_{x}+\hat{B}_{y}^{\dagger}\hat{B}_{y}+\hat{C}_{S}^{\dagger}\hat{C}_{S}+\hat{C}_{P}^{\dagger}\hat{C}_{P}\right).\nonumber \\
\label{eq:H0}
\end{eqnarray}
In this frame, in the presence of the probe beam the state $|\Psi (t)\rangle_I$ satisfies ($\hbar=1$)
\begin{eqnarray}
i\frac{d}{dt}|\Psi (t)\rangle_I=H_I|\Psi (t)\rangle_I,
\end{eqnarray}
where
\begin{eqnarray}
H_I& = & \delta_{p}\left(\hat{a}_{x}^{\dagger}\hat{a}_{x}+\hat{B}_{x}^{\dagger}\hat{B}_{x}+\hat{B}_{y}^{\dagger}\hat{B}_{y}+\hat{C}_{S}^{\dagger}\hat{C}_{S}+\hat{C}_{P}^{\dagger}\hat{C}_{P}\right),\nonumber \\
 &  & +\left(\frac{\Omega_{N}}{2}\hat{B}_{x}^{\dagger}\hat{a}_{x}+\frac{g_{\alpha}}{2}\hat{C}_{S}^{\dagger}\hat{B}_{y}+\frac{g_{\beta}}{2}\hat{C}_{S}^{\dagger}\hat{C}_{P}+\rm{h.c.}\right)\nonumber \\
 &  & -i\frac{\Delta}{2}\left(\hat{B}_{x}^{\dagger}\hat{B}_{y}-\hat{B}_{y}^{\dagger}\hat{B}_{x}\right)+i\sqrt{\frac{\kappa I_{p}}{2\omega_{p}}}\left(\hat{a}_{x}^{\dagger}-\hat{a}_{x}\right).\nonumber\\
\end{eqnarray}

\appendix


\begin{thebibliography}{99}
\bibitem{peak}In addition to the three peaks shown in Fig.~\ref{fig:PT}, in our system there are also two transmission peaks corresponding to the dressed states (bright states)  involving the
$^{3}\mathrm{S}_{1}$ state. These two peaks appear at quite large $\delta_{p}$, and thus are not shown in Fig.~\ref{fig:PT}.
\end{thebibliography}
\end{document}